\input harvmac.tex
 \input epsf.tex
 \input amssym

\def\figin{\epsfcheck\figin}\def\figins{\epsfcheck\figins}
\def\epsfcheck{\ifx\epsfbox\UnDeFiNeD
\message{(NO epsf.tex, FIGURES WILL BE IGNORED)}
\gdef\figin##1{\vskip2in}\gdef\figins##1{\hskip.5in}
\else\message{(FIGURES WILL BE INCLUDED)}%
\gdef\figin##1{##1}\gdef\figins##1{##1}\fi}
\def\DefWarn#1{}
\def\figinsert{\goodbreak\midinsert}
\def\ifig#1#2#3{\DefWarn#1\xdef#1{fig.~\the\figno}
\writedef{#1\leftbracket fig.\noexpand~\the\figno} %
\figinsert\figin{\centerline{#3}}\medskip\centerline{\vbox{\baselineskip12pt
\advance\hsize by -1truein\noindent\footnotefont{\bf
Fig.~\the\figno:} #2}}
\bigskip\endinsert\global\advance\figno by1}

\def\unit{\relax{\rm 1\kern-.26em I}}
\def\nada{\relax{\rm 0\kern-.30em l}}


\def \pa {\partial}

\def \eps {\epsilon}

\def\p{\partial}

\noblackbox
\def\IL{\relax{\rm I\kern-.18em L}}
\def\IH{\relax{\rm I\kern-.18em H}}
\def\IR{\relax{\rm I\kern-.18em R}}
\def\IC{\relax\hbox{$\inbar\kern-.3em{\rm C}$}}
\def\IZ{\relax\ifmmode\mathchoice
{\hbox{\cmss Z\kern-.4em Z}}{\hbox{\cmss Z\kern-.4em Z}} {\lower.9pt\hbox{\cmsss Z\kern-.4em Z}}
{\lower1.2pt\hbox{\cmsss Z\kern-.4em Z}}\else{\cmss Z\kern-.4em Z}\fi}

\def\CO {{\cal O}}

\def\CS {{\cal S}}
\def\CA{{\cal A}}


\def\CO {{\cal O}}

\def\CS {{\cal S }}

\def\diff{{\rm diff}}

\def\zb {\bar{z}}

\font\manual=manfnt \def\dbend{\lower3.5pt\hbox{\manual\char127}}

\def\ip{${\cal I}^+$}
\def\lf{{\cal L}_f}

\lref\StromingerJFA{
  A.~Strominger,
  ``On BMS Invariance of Gravitational Scattering,''
JHEP {\bf 1407}, 152 (2014).
[arXiv:1312.2229 [hep-th]].
}

\lref\BieriADA{
  L.~Bieri and D.~Garfinkle,
  ``A perturbative and gauge invariant treatment of gravitational wave memory,''
Phys.\ Rev.\ D {\bf 89}, 084039 (2014).
[arXiv:1312.6871 [gr-qc]].
}

\lref\TolishBKA{
  A.~Tolish and R.~M.~Wald,
  ``Retarded Fields of Null Particles and the Memory Effect,''
[arXiv:1401.5831 [gr-qc]].
}

\lref\TolishODA{
  A.~Tolish, L.~Bieri, D.~Garfinkle and R.~M.~Wald,
  ``Examination of a simple example of gravitational wave memory,''
Phys.\ Rev.\ D {\bf 90}, 044060 (2014).
[arXiv:1405.6396 [gr-qc]].
}

\lref\CardosoPA{
  V.~Cardoso, O.~J.~C.~Dias and J.~P.~S.~Lemos,
  ``Gravitational radiation in D-dimensional space-times,''
Phys.\ Rev.\ D {\bf 67}, 064026 (2003).
[hep-th/0212168].
}

\lref\HofmanAR{
  D.~M.~Hofman and J.~Maldacena,
  ``Conformal collider physics: Energy and charge correlations,''
JHEP {\bf 0805}, 012 (2008).
[arXiv:0803.1467 [hep-th]].
}
\lref\noteonsoft{
J. Maldacena and A. Zhiboedov, ``Notes on Soft Factors", unpublished (2012).
}

\lref\ZeldPoln{
Ya.~B.~Zeldovich and A.~G.~Polnarev, Sov. Astron. 18, 17 (1974)
}

\lref\BraginskyIA{
  V.~B.~Braginsky and L.~P.~Grishchuk,
  ``Kinematic Resonance and Memory Effect in Free Mass Gravitational Antennas,''
Sov.\ Phys.\ JETP {\bf 62}, 427 (1985), [Zh.\ Eksp.\ Teor.\ Fiz.\  {\bf 89}, 744 (1985)].
}

\lref\bragthorne{
V.~B.~Braginsky, K.~S.~Thorne,
"Gravitational-wave bursts with memory and experimental prospects."
Nature 327.6118 (1987): 123-125.
}

\lref\thorne{
K.~S.~Thorne,
"Gravitational-wave bursts with memory: The Christodoulou effect."
Physical Review D 45.2 (1992): 520.
}

\lref\ManasseZZ{
  F.~K.~Manasse and C.~W.~Misner,
  ``Fermi Normal Coordinates and Some Basic Concepts in Differential Geometry,''
J.\ Math.\ Phys.\  {\bf 4}, 735 (1963).
}

\lref\PoissonNH{
  E.~Poisson, A.~Pound and I.~Vega,
  ``The Motion of point particles in curved spacetime,''
Living Rev.\ Rel.\  {\bf 14}, 7 (2011).
[arXiv:1102.0529 [gr-qc]].
}

\lref\ChristodoulouCR{
  D.~Christodoulou,
  ``Nonlinear nature of gravitation and gravitational wave experiments,''
Phys.\ Rev.\ Lett.\  {\bf 67}, 1486 (1991).
}

\lref\BlanchetBR{
  L.~Blanchet and T.~Damour,
  ``Hereditary effects in gravitational radiation,''
Phys.\ Rev.\ D {\bf 46}, 4304 (1992).
}

\lref\BlanchetBR{
  L.~Blanchet and T.~Damour,
  ``Hereditary effects in gravitational radiation,''
Phys.\ Rev.\ D {\bf 46}, 4304 (1992).
}

\lref\TolishBKA{
  A.~Tolish and R.~M.~Wald,
  ``Retarded Fields of Null Particles and the Memory Effect,''
Phys.\ Rev.\ D {\bf 89}, no. 6, 064008 (2014).
[arXiv:1401.5831 [gr-qc]].
}

\lref\TolishODA{
  A.~Tolish, L.~Bieri, D.~Garfinkle and R.~M.~Wald,
  ``Examination of a simple example of gravitational wave memory,''
Phys.\ Rev.\ D {\bf 90}, 044060 (2014).
[arXiv:1405.6396 [gr-qc]].
}

\lref\AshtekarZSA{
  A.~Ashtekar,
  ``Geometry and Physics of Null Infinity,''
[arXiv:1409.1800 [gr-qc]].
}

\lref\CamanhoAPA{
  X.~O.~Camanho, J.~D.~Edelstein, J.~Maldacena and A.~Zhiboedov,
  ``Causality Constraints on Corrections to the Graviton Three-Point Coupling,''
[arXiv:1407.5597 [hep-th]].
}

\lref\BieriHQA{
  L.~Bieri and D.~Garfinkle,
  ``An electromagnetic analogue of gravitational wave memory,''
Class.\ Quant.\ Grav.\  {\bf 30}, 195009 (2013).
[arXiv:1307.5098 [gr-qc]].
}

\lref\HeLAA{
  T.~He, V.~Lysov, P.~Mitra and A.~Strominger,
  ``BMS supertranslations and Weinberg's soft graviton theorem,''
[arXiv:1401.7026 [hep-th]].
}

\lref\FavataZU{
  M.~Favata,
  ``The gravitational-wave memory effect,''
Class.\ Quant.\ Grav.\  {\bf 27}, 084036 (2010).
[arXiv:1003.3486 [gr-qc]].
}

\lref\LudvigsenKG{
  M.~Ludvigsen,
  ``Geodesic Deviation At Null Infinity And The Physical Effects Of Very Long Wave Gravitational Radiation,''
Gen.\ Rel.\ Grav.\  {\bf 21}, 1205 (1989)..
}

\lref\WinicourSKA{
  J.~Winicour,
  ``Global aspects of radiation memory,''
[arXiv:1407.0259 [gr-qc]].
}

\lref\SachsZZA{
  R.~Sachs,
  ``Asymptotic symmetries in gravitational theory,''
Phys.\ Rev.\  {\bf 128}, 2851 (1962)..
}

\lref\bms{H.~Bondi, M.~G.~J.~van der Burg and A.~W.~K.~Metzner,
  ``Gravitational waves in general relativity. 7. Waves from axisymmetric isolated systems,''
Proc.\ Roy.\ Soc.\ Lond.\ A {\bf 269}, 21 (1962);
  R.~K.~Sachs,
  ``Gravitational waves in general relativity. 8. Waves in asymptotically flat space-times,''
Proc.\ Roy.\ Soc.\ Lond.\ A {\bf 270}, 103 (1962)..
}

\lref\BarnichEB{
  G.~Barnich and C.~Troessaert,
JHEP {\bf 1005}, 062 (2010).
[arXiv:1001.1541 [hep-th]].
}

\lref\ManasseZZ{
  F.~K.~Manasse and C.~W.~Misner,
  ``Fermi Normal Coordinates and Some Basic Concepts in Differential Geometry,''
J.\ Math.\ Phys.\  {\bf 4}, 735 (1963)..
}

\lref\PoissonNH{
  E.~Poisson, A.~Pound and I.~Vega,
  ``The Motion of point particles in curved spacetime,''
Living Rev.\ Rel.\  {\bf 14}, 7 (2011).
[arXiv:1102.0529 [gr-qc]].
}

\lref\WeinbergNX{
  S.~Weinberg,
  ``Infrared photons and gravitons,''
Phys.\ Rev.\  {\bf 140}, B516 (1965)..
}

\lref\LudvigsenKG{
  M.~Ludvigsen,
  ``Geodesic Deviation At Null Infinity And The Physical Effects Of Very Long Wave Gravitational Radiation,''
Gen.\ Rel.\ Grav.\  {\bf 21}, 1205 (1989)..
}

\lref\MisnerQY{
  C.~W.~Misner, K.~S.~Thorne and J.~A.~Wheeler,
  ``Gravitation,''
San Francisco 1973, 1279p.
}

\lref\BieriHQA{
  L.~Bieri and D.~Garfinkle,
  ``An electromagnetic analogue of gravitational wave memory,''
Class.\ Quant.\ Grav.\  {\bf 30}, 195009 (2013).
[arXiv:1307.5098 [gr-qc]].
}

\lref\KinoshitaUR{
  T.~Kinoshita,
  ``Mass singularities of Feynman amplitudes,''
J.\ Math.\ Phys.\  {\bf 3}, 650 (1962)..
}

\lref\LeeIS{
  T.~D.~Lee and M.~Nauenberg,
  ``Degenerate Systems and Mass Singularities,''
Phys.\ Rev.\  {\bf 133}, B1549 (1964)..
}
\lref\WangZLS{
  J.~B.~Wang, G.~Hobbs, W.~Coles, R.~M.~Shannon, X.~J.~Zhu, D.~R.~Madison, M.~Kerr and V.~Ravi {\it et al.},
  ``Searching for gravitational wave memory bursts with the Parkes Pulsar Timing Array,''
[arXiv:1410.3323 [astro-ph.GA]].
}
\lref\HofmanAR{
  D.~M.~Hofman and J.~Maldacena,
  ``Conformal collider physics: Energy and charge correlations,''
JHEP {\bf 0805}, 012 (2008).
[arXiv:0803.1467 [hep-th]].
}
\lref\ck{
  D.~Christodoulou and S.~Klainerman,
  ``The Global nonlinear stability of the Minkowski space,''
Princeton University Press, Princeton, 1993.
}
\lref\yau{
  L.~Bieri, P.~Chen and S.~T.~Yau,
Class.\ Quant.\ Grav.\  {\bf 29}, 215003 (2012).
[arXiv:1110.0410 [astro-ph.CO]].
}
\lref\KulishUT{
  P.~P.~Kulish and L.~D.~Faddeev,
  ``Asymptotic conditions and infrared divergences in quantum electrodynamics,''
Theor.\ Math.\ Phys.\  {\bf 4}, 745 (1970), [Teor.\ Mat.\ Fiz.\  {\bf 4}, 153 (1970)]..
}

\lref\WareZJA{
  J.~Ware, R.~Saotome and R.~Akhoury,
  ``Construction of an asymptotic S matrix for perturbative quantum gravity,''
JHEP {\bf 1310}, 159 (2013).
[arXiv:1308.6285 [hep-th]].
}

\lref\WisemanSS{
  A.~G.~Wiseman and C.~M.~Will,
  ``Christodoulou's nonlinear gravitational wave memory: Evaluation in the quadrupole approximation,''
Phys.\ Rev.\ D {\bf 44}, 2945 (1991).
}

\lref\FavataZU{
  M.~Favata,
Class.\ Quant.\ Grav.\  {\bf 27}, 084036 (2010).
[arXiv:1003.3486 [gr-qc]].
}

%

\Title{\vbox{\baselineskip12pt}} {\vbox{
\centerline {Gravitational Memory, BMS Supertranslations} \centerline{ and Soft Theorems}}} \centerline{
Andrew Strominger and Alexander Zhiboedov} \vskip.1in \centerline{\it Center for the Fundamental Laws of Nature}\centerline{\it
Harvard University, Cambridge, MA 02138 USA}

\vskip.1in \centerline{\bf Abstract} { The transit of a gravitating radiation pulse past arrays of detectors stationed near future null infinity in the vacuum is considered. It is shown that the relative positions and clock times of the detectors before and after the radiation transit differ by a BMS supertranslation. An  explicit expression for the supertranslation in terms of moments of the radiation energy flux is given. The relative spatial displacement found for a pair of nearby detectors reproduces the well-known and potentially measurable  gravitational memory effect. The displacement memory formula is shown to be equivalent to Weinberg's formula for soft graviton production.
 }

\Date{}

 \listtoc\writetoc
\vskip 0.5in \noindent

\newsec{Introduction}

The passage of a finite pulse of radiation or other forms of energy through a region of spacetime produces a gravitational field which moves nearby detectors.  The final positions of a pair of nearby detectors are generically displaced relative to the initial ones according to a simple and universal formula \refs{\ZeldPoln\BraginskyIA\bragthorne\LudvigsenKG\ChristodoulouCR\WisemanSS\thorne\BlanchetBR\yau\TolishBKA-\TolishODA}. This effect is known as gravitational memory. Direct measurement of the gravitational memory effect may be possible in the coming decades, see $e.g.$ the recent work \refs{\WangZLS , \FavataZU}.

According to Bondi, Metzner, van der Burg and Sachs (BMS)\bms, the classical vacuum in general relativity is highly degenerate. The different vacua are related by the so-called `supertranslations', which are spontaneously broken `BMS symmetries'. In quantum language, these vacua differ by the addition of soft ($i.e$ zero-energy) gravitons.
In this paper we will show that the passage of radiation through a region induces a transition from one such vacuum to another. An explicit formula (involving moments of the radiation energy flux) is derived for the BMS supertranslation which
relates the initial and final vacua. Moreover, relative positions and clock times of a family of detectors stationed in the vacuum are shown to be related by the same supertranslation. This observation provides a concrete operational meaning to BMS transformations.

The relative spatial displacement  of nearby detectors following from the radiation-induced BMS transformation is precisely
the standard gravitational memory. We find that certain families of nearby detectors undergo, in addition to the standard spatial memory displacement, a relative time delay. It would be of interest to investigate potential experimental consequences.

Recently it has been shown \refs{\StromingerJFA} that the the observable consequences of BMS symmetry are embodied in the soft graviton scattering  amplitudes which they universally determine. Herein we show that  the Weinberg formula \WeinbergNX\ for soft graviton production is
essentially a rewriting of the formula for gravitational memory, establishing compatibility of \StromingerJFA\ with the current work. However, while it is quite difficult to imagine a real experiment which directly measures soft gravitons,  there is already a sizable literature on observation of gravitational memory. Hence the memory effect provides both a conceptually and observationally useful reformulation of BMS symmetry.

This paper is organized as follows. Section 2 establishes notation and briefly reviews BMS supertranslations. In section 3 we show that a finite  duration radiation pulse crossing null infinity can be viewed as a  domain wall mediating a transition between two inequivalent vacua of the gravitational field. Our central formula (3.7) is derived, involving the convolution of the radiative energy flux with a Green function, for the specific supertranslation which relates the initial and final vacua. Section 4 considers the effects of this transition on two types (inertial and fixed-angle) of detectors, and show that it can be understood as a supertranslation acting on the detector worldlines and
clocks.  This is shown to reproduce the  spatial gravitational memory effect. It further elucidates a clock desynchronization effect with potentially observable consequences for (fixed-angle) detectors.  In section 5 we show that the gravitational memory formula, in the form given by Braginsky and Thorne \bragthorne\ is, after a change of variables and notation, identical to Weinberg's soft graviton formula. In section 6 we point out that black holes are not invariant under supertranslations and therefore, in tension with the standard lore, carry an infinite amount of hair which encodes memories  of how they were formed.  Appendix A contains details on subleading corrections to large-radius geodesics. Appendix B demonstrates compatibility of our results with the interesting recent analyses by Tolish et. al. \refs{\TolishBKA,\TolishODA} of a specific example of the memory effect.

The existence of a  connection between gravitational memory and BMS symmetry is known and has been discussed periodically: see for example \refs{\AshtekarZSA, \WinicourSKA}. We expect the relation between asymptotic symmetries and memory to extend to other systems such as gauge theories. In particular in gauge theories the passage of charge through \ip\ should be remembered by angle-dependent gauge transformations on charged detectors.

\newsec{BMS review}

The metric of an asymptotically flat spacetime in retarded Bondi coordinates takes the asymptotic form
\eqn\asymptflat{\eqalign{
d s^2 &= - d u^2 - 2 du dr + 2 r^2  \gamma_{z \bar z} 
dz d \bar z \cr
&+2{m_B \over r} d u^2 + r C_{zz} d z^2 + r C_{\bar z \bar z} d \bar z^2 + D^{z} C_{z z} du dz + D^{\bar z} C_{\bar z \bar z} du d \bar z+...
}}
where $\gamma_{z\zb}={2 \over (1+z\zb)^2}$ is the unit metric on $S^2$, $D_z$ is the $\gamma$-covariant derivative and subleading terms are suppressed by powers of $r$.\foot{In particular we have corrections ${1 \over 4 r^2} C_{zz} C^{zz} du dr + \gamma_{z \bar z} C_{z z} C^{z z} dz d \bar z $ which contribute to the Einstein equations at the same order.} The Bondi mass aspect $m_B$ and $C_{zz}$ are related by the constraint equation $G_{u u} = 8 \pi G T_{u u}^{M}$ on \ip
\eqn\relation{\eqalign{
\pa_{u} m_{B} &={1 \over 4}  \left[D_z^2 N^{z z} + D^2_{\bar z} N^{\bar z \bar z} \right] - T_{uu}, \cr
T_{u u} &= {1 \over 4} N_{z z} N^{z z} + 4 \pi G \lim_{r \to \infty} [r^2 T^{M}_{u u}] .
}}
where $N_{zz}=\p_uC_{zz}$ is the Bondi news, $T^{M}$ is the matter stress tensor and $T_{uu}$ is the total energy flux through a given point on ${\cal I}^{+}$. The asymptotic form of the metric \asymptflat\  is preserved by infinitesimal supertranslations \bms\
\eqn\supertrl{\eqalign{
u &\to u-f,~~~~,r \to r -  D^{z} D_{z} f,\cr
z &\to z + {1 \over r} D^{ z} f,~~~~\zb \to\zb + {1 \over r} D^{\bar z} f, ~~~~~~~f=f(z,\zb),
}}
whose generating vector fields we denote
\eqn\supertransl{
\zeta_{f} = f \pa_{u} + D^{z} D_{z} f \pa_{r} - {1 \over r} (D^{\bar z} f \pa_{\bar z} + D^{z} f \pa_{z})  .
}
The Lie derivative action on the asymptotic data is
\eqn\sx{\eqalign{
\lf m_B&=f\pa_u m_B, \cr
~~~~\lf C_{zz}&=fN_{zz}-2D_z^2 f.
}}

According to BMS \bms\ two spacetimes related by supertranslations should be regarded as physically inequivalent.

\newsec{BMS vacuum transitions}

Consider spacetimes which, prior to some retarded time $u_i$ on  \ip,
are asymptotically well-approximated by  Schwarzschild with
\eqn\sr{m_B=M_i={\rm constant}, ~~~~C_{zz}=0,}
while for $u>u_f$ they are also nearly asymptotically Schwarzschild\foot{We exclude for simplicity cases with nonzero initial or final ADM momentum. }
\eqn\dk{m_B=M_f ={\rm constant}, ~~~~C_{zz}\neq 0.}
During the intermediate interval   $u_i<u<u_f$  the Bondi news and/or  total radiation flux $T_{uu}$  is nonzero on \ip.\foot{Generic spacetimes may have long time radiation tails outside this interval,  but for our purposes making the radiation flux outside the interval arbitrarily small is good enough.} Christodoulou and Klainerman \ck\ considered spacetimes of this type with $M_f=0$, where  $u_i$ and $u_f$ must be taken early and late enough to capture most of the long time tails. For nonzero $M_f$ the late time geometry could for example be a stable star or black hole.

\ifig\process{Remembrance of things passed. We consider a transit of radiation through a set of detectors in the vicinity of the future null infinity ${\cal I}^+$. Detectors are located at large $r_0$ and inserted at different points on the sphere $S^2$ separated by distance $L$. Change in the vacuum state is detected by the net displacement $\Delta L$.  The new vacuum is related to the old one by the supertranslation $C(z,\bar z)$.}{\epsfxsize3.5in \epsfbox{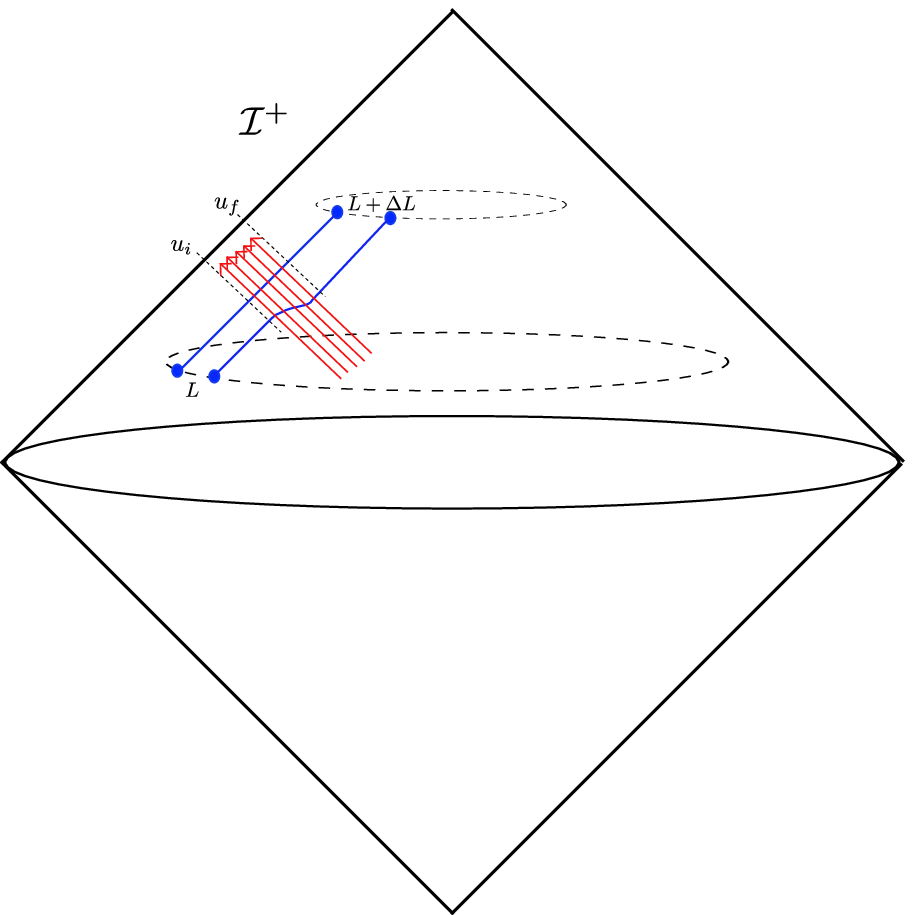} }

The initial and final regions of \ip\ before $u_i$ and after $u_f$ are in the vacuum in the sense that
$N_{zz}=0$: the radiative modes are unexcited.  According to BMS, the vacuum is not unique.
It is characterized by any $u$-independent $C_{zz}$ obeying
\eqn\gf{D_{\zb}^2 C_{zz } -D^2_{ z} C_{\bar z \bar z}=0.}
The general solution to this equation is
\eqn\bhj{C_{zz}=-2D_z^2C(z,\zb).}
Comparison  with \sx\ implies that the different vacua are related by supertranslations under which $C\to C+f$. The supertranslation which relates the initial and final vacua can be determined by integrating the constraint \relation\ over the transition interval  $u_i<u<u_f$.
Defining
\eqn\eed{ \Delta C_{zz}=C_{zz}(u_f)-C_{zz}(u_i),~~~~\Delta m_B=M_f-M_i,}
and using \gf\ one finds
\eqn\resultint{
D_z^2 \Delta C^{z z} = 2\int_{u_i}^{u_f} d u \  T_{u u}+ 2\Delta m_{B} .
}
Note that the second term just subtracts the constant zero mode of the first.\foot{Had we allowed for net momentum loss as well as energy loss the right hand side would also contain a term subtracting the angular momentum $\ell=1$ mode.}
The supertranslation $\Delta C$ which produces such a $\Delta C_{zz}$ is obtained by
inverting $D_z^2D_{\zb}^2$:
\eqn\frf{
\Delta C(z,\zb)=2 \int d^2 z' \gamma_{z' \bar z'} G(z,\zb;z',\zb')\left( \int_{u_i}^{u_f} d u \ T_{u u} (z',\zb')  + \Delta m_{B} \right)
}
where
\eqn\manifest{\eqalign{
G(z,\zb;z',\zb') &=- {1 \over \pi}\sin^2 {\Theta \over 2}  \log\sin^2 { \Theta \over 2} , ~~~ \sin^2 {\Theta(z,z') \over 2} \equiv{|z-z'|^2\over (1 + z' \bar z ')(1 + z \bar z) },  \cr
D_z^2D_{\zb}^2G(z,\zb;z',\zb')&=- \gamma_{z \bar z} \delta^2(z-z') +\cdots.
}}
 If we plug \manifest\ into \frf\ and act with $D_z^2D_{\zb}^2$ using $\pa_z \pa_{\bar z} \log |z - z_i|^2 = 2 \pi \ \delta^{(2)}(z - z_i)$ the delta function piece produces the RHS of \resultint\ while the remaining  terms integrate to zero due to the energy-momentum conservation. $C(z,\bar z)$ is unique up to the 4 global spacetime translations
\eqn\conservation{
f_{global} = c_0 + {c_1 (1 - z \bar z) + c_2 z  + c_3 \bar z +\bar c_3 z \over 1 + z \bar z}
}
which do not affect $C_{zz}$.

This discussion could be generalized to allow for initial and final momentum, or multiple vacuum transitions induced by multiple well-separated radiation intervals.

To summarize, the passage of radiation through \ip\ changes the vacuum by a  BMS transformation. The BMS transformation relating the initial and final vacuum is given in \frf\ by an integral of the total radiation flux over the transition interval.

\newsec{Gravitational memory}

In this section we will relate the BMS transformation of the vacuum to the gravitational memory effect.
Towards this end we introduce two families of observers or detectors at large $r$. The
first, which we refer to as BMS (or fixed-angle) detectors, travel along worldlines at fixed radius and angle:
\eqn\bmsd{X^\mu_{BMS}(s)=(s,r_0,z_0,\zb_0),}
where $r_0$ is large.  The assertion that BMS diffeomporhisms are physically nontrivial is equivalent to the statement that it is meaningful to discuss observations at a fixed value of $z$ near \ip .  Such observations are convenient as they behave simply under the action of BMS.  The second family of detectors are inertial ones  moving along  geodesics
\eqn\ewq{\pa_s^2X^\mu_{geo}(s)+\Gamma^\mu_{\nu\lambda}\pa_sX_{geo}^\nu(s)X_{geo}^\lambda(s)=0.}
At large $r_0$ the BMS detectors are nearly inertial. One may readily check (see Appendix A) that
\eqn\ret{ X^{u,r}_{BMS}(s)= X^{u,r}_{geo}( s)+\CO({1 \over r_0}),~~~~~~~X^{z}_{BMS}( s)= X^{z}_{geo}( s)+\CO({1 \over r_0^2}).}
 The truly inertial detectors however do not remain at fixed $r$ or $z$, so over a long period of time $u>r_0$ the radius can become small. Hence we must consider only retarded time lapses which are parametrically less than $r_0$.

The relevant type of detector  - BMS or inertial - depends on the  application in question. For example the eLisa
detectors move on geodesic orbits and so are perhaps best modeled by inertial detectors. On the other hand the LIGO detector are at fixed separations on the earth and are not geodesic. It would be interesting to understand what type of detector array is well-approximated by BMS detectors.

\subsec{BMS detector memory}

Let us now consider what happens to the BMS detector worldlines in the setup of the previous section when they encounter a pulse of  radiation passing to \ip.  Let us denote the initial positions of a pair of nearby detectors, detector 1 and detector 2, by   $z_{1}$ and $z_2$. They  are initially separated by a finite distance
\eqn\ssz{
L={2 r_0|\delta z|\over 1+z_1\zb_1},~~~~\delta z\equiv z_1-z_2
}
where we take $\delta z$ to be order $1 \over r_0$ and subleading corrections to $L$ are suppressed. As
$z_{1,2}$ are fixed in \bmsd, but the metric undergoes a transition described by \eed,  the radiation induces a change in the proper distance between the detectors. Computing the new distance between $z_1$ and $z_2$ using the metric \asymptflat\  gives
\eqn\rij{
\Delta L= {r_0 \over 2L}\Delta C_{zz}(z_1,\zb_1)\delta z^2+c.c.={(1+z_1\zb_1)^2 \over 8 } {L \over r_0} \left( \Delta C_{zz}(z_1,\zb_1) {\delta z \over \delta \zb }+c.c \right),
}
where $\Delta C_{zz}(z_1,\zb_1)$ is given according to \eed\ in terms of the energy flux as
\eqn\christmemform{
\Delta C_{z z}(z,\bar z) = {4 \over \pi } \int d^2 z' \gamma_{z' \bar z'} {\bar z - \bar z' \over z - z'}{(1+ z' \bar z)^2 \over (1 + z' \bar z')(1 + z \bar z)^3}\left( \int_{u_i}^{u_f} d u \ T_{u u} (z',\zb')  + \Delta m_{B} \right)
}
This is precisely the standard formula for gravitational memory \refs{\ZeldPoln,\LudvigsenKG,\ChristodoulouCR , \BlanchetBR} .

Not only will the distances between BMS detectors be shifted, but if they are equipped with initially synchronized clocks they will no longer be synchronized after passage of the radiation.  This can be checked by sending a light ray from detector 1 to detector 2, stamping it with the time at detector 2 and then returning it to detector 1. If the clocks remain synchronized, the time stamp from detector 2 will be exactly midway between the light emission and reception times at detector 1.    A light ray emitted from $z_1$ will travel to $z_2$ in a retarded time interval $\delta_{12} u$
obeying
\eqn\oop{{r_0^2 }\gamma_{z\zb}\delta z\delta \zb + r_0\Delta C_{zz}\delta z\delta z + D^z\Delta C_{zz}\delta_{12} u \delta z -{1 \over 2} (\delta_{12} u)^2+c.c.=0.}
On the other hand, on the return trip, the change in $z$  has the opposite sign so the retarded time interval $\delta_{21} u$
obeys
\eqn\oops{{r_0^2 } \gamma_{z\zb}\delta z\delta \zb +r_0\Delta C_{zz}\delta z\delta z -D^z\Delta C_{zz}\delta_{21} u \delta z -{1 \over 2} (\delta_{21} u)^2+c.c.=0.}
The difference is\foot{The total elapsed time is, to leading order in $r_0$,
$\delta_{12} u+\delta_{21} u=2L.$}
\eqn\diff{ \delta_{12} u-\delta_{21} u=D^z\Delta C_{zz}\delta z +c.c..}
Since this is nonzero the clocks are not synchronized.

An alternate way of computing the memory and clock desynchronization is as follows. The proper distance and time delay observed in the above mentioned experiments is invariant under all diffeomorphisms, including BMS transformations. We may therefore eliminate all $\Delta C_{zz}$ terms in the late time metric by the inverse of the BMS transformation \frf\ which by construction obeys
\eqn\iop{2D_z^2f=\Delta C_{zz},}
so that $f=-\Delta C$
This will have the effect of resetting all the clocks and relabeling the positions of the family of BMS observers by \supertrl.
To see that this agrees with the previous analysis let  $\zeta = \delta z\pa_z+ \delta \bar z\pa_{\zb}$ denote the initial separation vector between the detectors. Using  \supertransl\ the action of the supertranslation \supertrl\  on this  separation  is
\eqn\separationch{\eqalign{
{\cal L}_f \zeta &=-\half\left( D_{z} f \delta z +  D_{\bar z} f \delta \bar z \right)\pa_u -\half\left( D_{z}^2 D^{z} f \delta z +  D_{\bar z}^2 D^{\bar z} f \delta \bar z \right)\pa_r\cr &~~+\left( {\gamma^{z \bar z} \over r} D_{\bar z}^2 f \delta \bar z + {1 + z \bar z \over 2 r} [2 \bar z D_{\bar z} f + (1+ z \bar z) D_{z} D_{\bar z} f] \delta z \right)\pa_z+c.c.. 
}}
To compare to the original coordinate system we evaluate the norm of the vector at $(z + {1 \over r} D^{ z} f,\zb + {1 \over r} D^{\bar z} f)$. The proper distance changes by
\eqn\rsij{
\Delta L={r_0\over 2L} \Delta C_{zz}(z_1,\zb_1)\delta z^2+c.c.
}
which agrees, as it must,  with \rij. The extra terms that appear in \separationch\ cancel against the change of the metric of the flat space evaluated at the shifted point.

To compare the time delay of the two detectors  two effects must be taken into account. First the transformation of $u\to u - f$ resets the clocks by  a relative amount $D_z f\delta z+c.c.$. A second effect arises because the relative radius changes by
\eqn\ws{ \delta r=-D_z^2D^zf\delta z+c.c.}
Due to the presence of the term $2dudr$ in the metric, this implies a difference proportional to $\delta r$ in the time lapses for light rays traveling from detector 1 to detector 2 and the reverse.  Adding these two effects, and using
\eqn\red{[D_z,D_{\zb}]D^{\zb}f=-D_zf,}
one finds
\eqn\tdiff{ \delta_{12} u-\delta_{21} u=D^z\Delta C_{zz}\delta z +c.c.,}
as expected.

In conclusion the effects of a radiation pulse passing through \ip\ on a family of BMS observers is characterized by
the induced supertranslation \frf. They may be equivalently described as leaving the worldlines unchanged and supertranslating the metric, or leaving the metric unchanged and supertranslating the observers. In either case they imply the familiar gravitational memory effect as well as clock desynchronization.

\subsec{Inertial detector memory}
Most discussions of gravitational memory involve inertial (rather than BMS) detectors moving on geodesics \ewq\ that are nearly, but not exactly, worldlines of constant $(r,z)$ and varying $u$.
According to \ret , the difference between the two worldlines is suppressed by powers of $r$. It immediately follows that the spatial gravitational memory formula \rij \ applies equally at large $r$ to either BMS or inertial detectors.

The situation is more subtle for the relative time delay. In that case, we found above that there
are two contributions which cancel at leading order, and the final result \tdiff\ is the sum of the subleading terms for each contribution. These subleading terms are in fact sensitive to the difference
between the BMS and inertial worldlines.
Direct computation reveals that, for inertial observers, the relative time delay actually vanishes at the order \tdiff\ , as we show in  appendix A. In Bondi coordinates, this cancelation looks miraculous. However it is in fact a consequence of the equivalence principle, which implies the existence of Fermi normal coordinates in which the connection vanishes everywhere along the worldlines of two neighboring geodesics. It follows there can be no discrepancy of order $L$ in the proper times and \tdiff\ hence must be cancelled by subleading geodesic corrections.

\newsec{Memory and soft theorems}

Recently it has been shown \refs{\StromingerJFA,\HeLAA} that Weinberg's soft graviton theorem \WeinbergNX\ is equivalent to - or more precisely is the Ward identity of - BMS invariance of the quantum gravity $\CS$-matrix.  In the preceding we have seen that the gravitational memory effect captures the consequences of BMS symmetry. In this section we show how  to directly understand the relation between the memory effect and the soft theorem without an interpolating  discussion of BMS symmetry.

Weinberg's soft graviton theorem \WeinbergNX\ is a universal relation between ($n\to m+1$)-particle with one final soft graviton and ($n\to m$)-particle  quantum field theory scattering amplitudes given by
\eqn\qet{
\lim_{\omega \to 0}\CA_{m+n+1}\bigl(p_1,...p_n; p'_1,...p'_m, (\omega k,\epsilon_{\mu\nu})\bigr)=\sqrt{8 \pi G} S_{\mu\nu} \eps^{\mu\nu} \CA_{m+n}\bigl(p_1,...p_n; p'_1,...p'_m\bigr) +\CO(\omega^0),}
where
\eqn\rft{
S_{\mu\nu}=\left(\sum_{j=1}^m{p_{j\mu}p_{j\nu} \over \omega k\cdot p_j} - \sum_{j=1}^n{p'_{j\mu}p'_{j\nu} \over \omega k \cdot p'_j}\right)^{TT}.
}
 In this expression $k=(\omega,\omega \vec  k)$ with $\vec k^2=1$ is the four-momentum and $\epsilon_{\mu\nu}$ the  transverse-traceless polarization tensor of the  graviton.  The superscript $TT$ denotes the transverse-traceless projection (as detailed in \MisnerQY) and $\mu,\nu$ indices refer to asymptotically Minkowskian coordinates with flat metric $\eta_{\mu\nu}$.

Here  we explicate the relation between memory and soft theorems in the general context considered by
Braginsky and Thorne \bragthorne. They analyzed the possible detection of ``burst memory waves" produced by the collision and scattering of large massive objects such as stars or black holes. They found that such collisions resulted in a net difference in the transverse traceless part of the asymptotic metric at \ip\ given by\foot{This is equation (1) of \bragthorne\  written with the normalization $h_{\mu\nu}\equiv {1 \over \sqrt {32 \pi G}}(g_{\mu\nu}-\eta_{\mu\nu})$, in the mostly plus $(-+...+)$ signature and in covariant gauge. }  \eqn\bth{
\Delta h_{\mu\nu}^{TT}(\vec k)={1 \over r_0} \sqrt{G \over 2 \pi}\left( \sum_{j=1}^n{p'_{j\mu}p'_{j\nu} \over \omega k \cdot p'_j} - \sum_{j=1}^m{p_{j\mu}p_{j\nu} \over \omega k\cdot p_j} \right)^{TT}.
}
Here we have $n$ ($m$) incoming (outgoing) objects with asymptotic momenta $p_{j\mu}$ ($p'_{j\mu}$). $k=(1,\vec k)$  is the null vector pointing  from the collision region to null infinity, and serves as a coordinate on the $S^2$ at \ip. Equation \bth\ was derived by solving the linearized Einstein equation with a retarded propagator. The gravitational memory of the collision is then simply constructed from \bth\ via \rij.

Evidently there are strong similarities between \bth\ and \rft. To make it more manifest we note the Fourier transform of $h^{TT}_{\mu\nu }(\omega,\vec k)$   on \ip\ can be written, using the stationary phase approximation at large $r$ \refs{\noteonsoft,\HeLAA}
\eqn\fourier{
h^{TT}_{\mu\nu}(\omega,\vec k) =
 4 \pi i \lim_{r \to \infty} r \int d u \ e^{i \omega u} h^{TT}_{\mu \nu} (u, r \vec k ) ,}
Assuming that $h^{TT}_{\mu\nu}(u,r\vec k)$ approaches finite but different values at $u\to \pm \infty$ and large $r=r_0$ it then follows\foot{ In the formulas above we assume that $\omega r \gg 1$ when taking the limits.} that \bth\ is proportional to the coefficient of the pole in $\omega$
\eqn\relation{
\Delta h^{TT}_{\mu \nu}(\vec k) ={1 \over 4 \pi i r_0} \lim_{\omega \to 0} \left( - i \omega h^{TT}_{\mu \nu}(\omega,\vec k) \right) .
}

Next we note that to linear order, the expectation value of the asymptotic metric fluctuation produced in the process of $n\to m$ scattering obeys
\eqn\frrr{\eqalign{\lim_{\omega \to 0} \omega h^{TT}_{\mu\nu }(\omega,k)\epsilon^{\mu \nu}&= \lim_{\omega \to 0}{ \omega \CA_{m+n+1}\bigl(p_1,...p_n; p'_1,...p'_m, (\omega k ,\epsilon_{\mu\nu})\bigr)\over\CA_{m+n}\bigl(p_1,...p_n; p'_1,...p'_m\bigr)} \cr &= \sqrt{8 \pi G} \epsilon^{\mu \nu}\lim_{\omega \to 0}\omega S_{\mu\nu}(\omega k)\cr &= \sqrt{8 \pi G}  \epsilon^{\mu \nu}\left( \sum_{j=1}^m{p_{j\mu}p_{j\nu} \over k\cdot p_j} - \sum_{j=1}^n{p'_{j\mu}p'_{j\nu} \over k\cdot p'_j} \right)^{TT}.}}
Inserting this into \relation\ we then see that this is equivalent as claimed to the Braginsky-Thorne result \bth.

\newsec{Measuring black hole hair}

One often hears that black holes have no hair. This statement does not take into account the subtleties associated with asymptotic structure at $\CI$. In particular, as we discussed in section 3,  a supertranslation maps the Schwarzschild solution to a physically inequivalent configuration. Hence black holes have a lush infinite head of supertranslation hair. This may bear on the information puzzle.

\ifig\bhhair{ Supertranslation hair of black holes. We consider formation of a black hole by infalling matter. During the process radiation is necessarily created and leaks to future null infinity, where it mediates the transition to a new vacuum state. When the black hole settles down the state at ${\cal I}^+$ is characterized by the supertranslation hair $C(z, \bar z)$. }{\epsfxsize2.0in \epsfbox{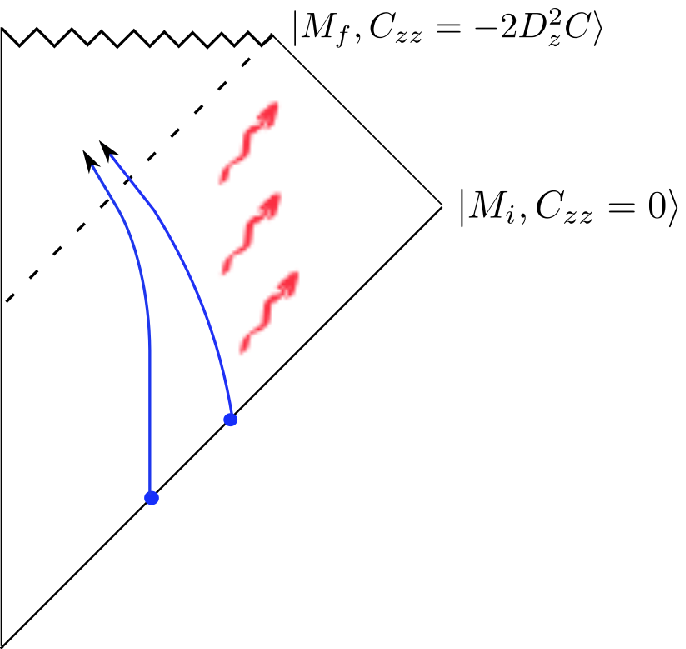} }

The present discussion clarifies the nature of supertranslation hair and how it can be measured classically. Let us consider, as a special case of the Braginsky-Thorne construction,  $N$ incoming stars which collide and collapse into a black hole. We station an array of evenly-spaced  detectors near future null infinity. The relative positions of these detectors will shift due to the memory effect as given in \bth\ and \rij. Hence the detector positions can record an infinite amount of data about how the black hole was formed. Black hole formed by different initial star configurations will carry different supertranslation hair.

\centerline{\bf Acknowledgements}
We are grateful to S. Gralla, D. Kapec, J. Maldacena, P. Mitra, S. Pasterski and A. Porfyriadis  for useful conversations. This work was supported in part by NSF grant 1205550.

\appendix{A}{Corrections to geodesics and Fermi coordinates}

Let us consider an asymptotic time-like observer with the velocity $v^{\mu}$ in the Bondi retarded coordinates. Its four-velocity is given by
\eqn\observer{
v^{\mu} = (1  , {m_{B}(u,z,\bar z) \over r}, - {1 \over 8} {(1 + z \bar z)^4 \over r^2} D_{z} C_{\bar z \bar z}(u, z , \bar z) ,  - {1 \over 8} {(1 + z \bar z)^4 \over r^2} D_{\bar z} C_{z z}(u, z, \bar z) ) .
}

Notice that for the first two components we work to the ${1  \over r}$ order and for the last two to the ${1 \over r^2}$. The reason for this will become clear below. This four-vector describes a geodesic $v^{\mu} \nabla_{\mu} v^{u,r} = O({1 \over r^2})$, $v^{\mu} \nabla_{\mu} v^{z, \bar z} = O({1 \over r^3})$. For the norm we have $v^{\mu} v_{\mu} = - 1 + O({1 \over r^2})$.

Equally well (by choosing different initial conditions) we could have chosen the four-velocity to be
\eqn\observerMore{\eqalign{
v^{\mu} &= (1 + { m_B(u_0, z ,\bar z) \over r} , {m_{B}(u,z,\bar z) - m_B(u_0, z ,\bar z) \over r}, \cr
&- {1 \over 8} {(1 + z \bar z)^4 \over r^2} D_{z} [C_{\bar z \bar z}(u, z , \bar z) - C_{\bar z \bar z}(u_0, z , \bar z)],  - {1 \over 8} {(1 + z \bar z)^4 \over r^2} [D_{\bar z} C_{z z}(u, z, \bar z) - D_{\bar z} C_{z z}(u_0, z, \bar z)] ) .
}}
In \observer\ we set $u$-independent initial values to zero. We do the same thing below since we are interested in the vacuum-to-vacuum transitions described in the bulk of the paper.

If we integrate over $u$ for a long enough time the corrections are not small since the correspondent integrals diverge. Below we always assume $r$ to be large enough (and the measurement time to be small enough) so that the corrections are small.

We also consider the orthonormal spatial basis
\eqn\spatial{\eqalign{
n^{\mu} &= (-1 , 1 - {m_{B} \over r},   {1 \over 8} {(1 + z \bar z)^4 \over r^2} D_{z} C_{\bar z \bar z},   {1 \over 8} {(1 + z \bar z)^4 \over r^2} D_{\bar z} C_{z z}), \cr
m^{\mu} &=(0,0, {1 \over r} {1+z \bar z \over 2} , - {1 \over r} {1+z \bar z \over 2}{(1 + z \bar z)^2  C_{z z} \over 4 r} ), \cr
\bar m^{\mu} &=(0,0,- {1 \over r} {1+z \bar z \over 2}{(1 + z \bar z)^2  C_{\bar z \bar z} \over 4 r},  {1 \over r} {1+z \bar z \over 2}  ) .
}}

All of these are parallel transported along $v^{\mu}$ to leading order in ${1 \over r}$ so that we have $v^{\mu} \nabla_{\mu} e_{i}^{\nu} = O({1 \over r^2})$ and are orthogonal to $v^{\mu}$, namely $v_{\mu}e_{i}^{\mu} = O({1 \over r^2})$. They are also normalized in the usual way $n.n = 1 + O({1 \over r^2})$, $m . \bar m = {1 \over 2} + O({1 \over r^2})$, $n . m = n . \bar m = m .m = \bar m . \bar m = O({1 \over r^2})$. Physically, these vectors describe a set of gyroscopes that is carried by an observer.

Based on this we can introduce Fermi normal coordinates which are the coordinates that describe physics that that the observer experiences in the vicinity of his location. Namely introducing $e^{\mu}_{0} = v^{\mu}$, $e^{\mu}_{3} = n^{\mu}$, $e^{\mu} = m^{\mu}$, $\bar e^{\mu} = \bar m^{\mu}$ we introduce corresponding coordinates $x_i^{\mu}$ such that the metric takes the form
\eqn\metric{
d s^2 = - d x_0^2 + d x_3^2 + d x d \bar x + O(x^2)
}
and the leading corrections are related to $R_{a b c d} = R_{\mu \nu \rho \sigma} e^{\mu}_{a} e^{\nu}_{b} e^{\rho}_{c} e^{\sigma}_{d}$ \refs{ \ManasseZZ, \PoissonNH } where the Riemann tensor is evaluated along the geodesic $\gamma(\tau)$. The equations for geodesics take the form
\eqn\deviation{
\ddot x^{i} = R^{i}_{\ 00 j} x^{j}.
}
For the case in hand the only contribution at the ${1 \over r}$ order comes from $R_{x 00 x} = {(1 + z \bar z)^2 \over 8 r} \pa_u^2 C_{z z }$ and $R_{\bar x 00 \bar x} = {(1 + z \bar z)^2 \over 8 r} \pa_u^2 C_{\bar z \bar z }$ which describe the ordinary gravitational memory. The same analysis was done in \LudvigsenKG .

It is clear in Fermi coordinates that there is no relative time shift linear in $x$ unless the original observer is accelerated. This is the case for BMS observers and it is the source of time desynchronization linear order in $x$.

We can also analyze the nearby geodesic directly in Bondi coordinates as explained in the main body of the paper. The result of course does not depend on the coordinate system we used.

\appendix{B}{Massive particle decay}

It is instructive to compare our results to those of Tolish et. el.  \TolishBKA , \TolishODA . The starting point of their work is the geodesic deviation equation for a small perturbation around the flat space
\eqn\geoddev{\eqalign{
{d^2 D_\mu \over d t^2} &= - R_{t \mu t \nu} D^{\nu}, \cr
\Delta D_{\mu} &= M_{\mu \nu} D^{\nu}.
}}
For the cases considered in \TolishODA,  symmetries imply that
\eqn\symmriem{
R_{t \mu t \nu} = W (\theta_\mu \theta_\nu - \phi_\mu \phi_\nu)
}
where $\theta^\mu$ and $\phi^\mu$ are the unit vector fields on the sphere.

In the Bondi coordinates we have for the tensor $\hat M_{\mu \nu} = \theta_\mu \theta_\nu - \phi_\mu \phi_\nu $
\eqn\newtensor{
\hat M_{z z} = {2 \over (1 + z \bar z)^2} {\bar z \over z},~~~ \hat M_{\bar z \bar z} = {2 \over (1 + z \bar z)^2} {z \over \bar z} .
}

Consider now the following situation. A particle at rest of mass $M$ decays into a massless particle with energy $E$ moving in the $\hat z$-direction and the particle of mass $M'$ moving in the $-\hat z$-direction. On the sphere it corresponds to $z = \bar z = 0$ for $\hat z$ and $z = \bar z = \infty$ for $- \hat z$.
The contribution of the massless particle is \TolishODA
\eqn\massless{
M_{\mu \nu} = {E \over r} (1 + \cos \theta) \hat M_{\mu \nu} = {E \over r} {2 \over 1 +  z \bar z} \hat M_{\mu \nu}
}
whereas for the massive one \TolishODA
\eqn\massive{
M_{\mu \nu} = {E^2 \over M r} {\sin \theta^2 \over 1 - {E \over M} \left( 1 - \cos \theta \right)} \hat M_{\mu \nu}  = {1 \over r} {|\vec p|^2 \sin \theta^2 \over p^0 + |\vec p | \cos \theta}  \hat M_{\mu \nu}
}
where $p^0 = M - E$, $|\vec p| = E$ by energy-momentum conservation. In the formulae above we set $G = 1$.

We now would like to reproduce the same formulas using the soft theorem which states that the memory is given by the soft factor\foot{The prescription for taking transverse-traceless part is thoroughly reviewed in \MisnerQY
\eqn\prescrtt{\eqalign{
h_{\mu \nu}^{TT} &= h_{\mu \nu} - n_\mu h_{\nu \lambda} n^\lambda - n_\nu h_{\mu \lambda} n^\lambda + n_\mu n_\nu (h_{\lambda d} n^\lambda n^d) \cr
&- {1 \over 2} (\delta_{\mu \nu} - n_\mu n_\nu) h_{\lambda d}\left( \delta^{\lambda d} - n^\lambda n^d \right).
}}
}
\eqn\softmem{
M_{\mu \nu} ={1 \over 2} \sqrt{32 \pi G} h_{\mu \nu}^{T T} =  {2 G \over r } \left( \sum_{i} {p^{i}_{\mu} p^{i}_{\nu} \over (p^{i} . n) } - \sum_{f} {p^{f}_{\mu} p^{f}_{\nu} \over (p^{f} . n) }\right)^{TT} ,
}
where we adopted field theoretical normalization $h_{\mu \nu} = {1 \over \sqrt{32 \pi G}} \left( g_{\mu \nu} - \eta_{\mu \nu} \right)$ and $n=(1,\vec n)$ is the unit four-vector in the direction of observation.
Plugging the momenta in the formula above reproduces the result of \TolishODA. For massless particles in the $(z, \bar z)$ coordinates it becomes
\eqn\softmemb{
M_{z z} =  {4 G \over r } \sum_{i} E_i {\bar z_i - \bar z \over z_i - z} {(1 + z_i \bar z)^2 \over (1 + z_i \bar z_i) (1+ z \bar z)^3 }.
}
In the example above we have $z_i = \bar z_i = 0$. Notice also that the $(z,\bar z)$-dependent kernel that appeared in \softmemb\ is identical to the one that appeared in \christmemform . Indeed, as pointed out in \thorne\ the Christodoulou memory effect can be thought as a generalization of the usual soft factor where instead of a finite set of particles approaching infinity we imagine arbitrary energy flux of gravitational radiation.
Any fixed energy scattering can produce at most finite number of massive particles. It means that a generic final state can be thought as the finite number of massive particles plus arbitrary complicated profile of radiation. The memory due to the radiation is captured by \christmemform , whereas for massive particles the contribution to the memory is simply given by \softmem .

\listrefs

\bye